\documentclass{aa}  

\usepackage{graphicx}
\usepackage{txfonts}
\usepackage{color}
\usepackage{stfloats}
\usepackage{multirow}
\usepackage{mhchem}
\usepackage{booktabs}

\newcommand{\cp}{Chem. Phys.}
\newcommand{\cpl}{Chem. Phys. Lett.}
\newcommand{\chemrev}{Chem. Rev.}

\newcommand{\ijms}{Int. J. Mass Spectrom.}

\newcommand{\jms}{J. Mol. Spectr.}

\newcommand{\jpc}{J. Phys. Chem.}
\newcommand{\jpcl}{J. Phys. Chem. Lett.}
\newcommand{\jpca}{J. Phys. Chem. A}

\newcommand{\natastro}{Nat. Astron.}

\newcommand{\pnas}{PNAS}

\newcommand{\science}{Science}

\begin{document}



\title{A search for the three isomers of cyano-1,3-butadiene in \mbox{TMC-1}:\\ Implications for bottom-up routes involving 1,3-butadiene\thanks{Based on observations carried out with the Yebes 40m telescope (projects 19A003, 20A014, 20D023, 21A011, 21D005, and 23A024). The 40m radio telescope at Yebes Observatory is operated by the Spanish Geographic Institute (IGN; Ministerio de Transportes, Movilidad y Agenda Urbana).}}

\titlerunning{Search for cyano-1,3-butadiene in \mbox{TMC-1}}
\authorrunning{Ag\'undez et al.}

\author{M.~Ag\'undez\inst{1}, C.~Cabezas\inst{1}, N.~Marcelino\inst{2,3}, B.~Tercero\inst{2,3}, R.~Fuentetaja\inst{1}, P.~de~Vicente\inst{3}, \and J.~Cernicharo\inst{1}}

\institute{
Instituto de F\'isica Fundamental, CSIC, Calle Serrano 123, E-28006 Madrid, Spain\\ \email{marcelino.agundez@csic.es, jose.cernicharo@csic.es} \and
Observatorio Astron\'omico Nacional, IGN, Calle Alfonso XII 3, E-28014 Madrid, Spain \and
Observatorio de Yebes, IGN, Cerro de la Palera s/n, E-19141 Yebes, Guadalajara, Spain
}

\date{Received; accepted}

 
\abstract
{The molecule 1,3-butadiene (CH$_2$CHCHCH$_2$) could play a key role in the synthesis of the cyclic molecules cyclopentadiene and benzene in cold dense clouds. Since 1,3-butadiene is non-polar, we searched for its cyano derivative, which exists in the form of three different polar isomers, in the cold dense cloud \mbox{TMC-1}. We used the most recent data obtained with the Yebes\,40m telescope in the Q band (31.0-50.3 GHz) in the frame of the QUIJOTE project. We do not detect any of the two isomers of 1-cyano-1,3-butadiene, and derive 3\,$\sigma$ upper limits to their column densities of 1.2\,$\times$\,10$^{10}$ cm$^{-2}$ and 2.0\,$\times$\,10$^{10}$ cm$^{-2}$ for $E$- and $Z$-1-cyano-1,3-butadiene, respectively. Our results are not consistent with those from \cite{Cooke2023}, who determine a column density of 3.8\,$\times$\,10$^{10}$ cm$^{-2}$ for $E$-1-cyano-1,3-butadiene in \mbox{TMC-1} using GBT data and a line stack technique. At the current level of sensitivity of our data, there is tentative evidence for the presence of the third cyano derivative isomer, 2-cyano-1,3-butadiene, although a firm detection must await more sensitive data. We derive an upper limit to its column density of 3.1\,$\times$\,10$^{10}$ cm$^{-2}$. This isomer cannot be formed in the reaction between CN and 1,3-butadiene, according to experimental and theoretical studies, and thus we speculate whether it could arise from neutral-neutral reactions like C$_2$H$_3$ + CH$_2$CHCN and CH$_2$CCN + C$_2$H$_4$. From the upper limit on the abundance of 1-cyano-1,3-butadiene derived here, we estimate that the abundance of 1,3-butadiene in \mbox{TMC-1} is below 10$^{-11}$-10$^{-10}$ relative to H$_2$. The low abundance inferred for 1,3-butadiene makes it unlikely that it plays an important role in bottom-up routes to cyclopentadiene and benzene.}

\keywords{astrochemistry -- line: identification -- ISM: individual objects (\mbox{TMC-1}) -- ISM: molecules -- radio lines: ISM}

\maketitle

\section{Introduction}

The hydrocarbon 1,3-butadiene (CH$_2$CHCHCH$_2$) is a very interesting molecule from an interstellar perspective because it shows a great reactivity with small hydrocarbon radicals, resulting in cyclization to form simple rings such as cyclopentadiene and benzene. The interest on the formation of these cycles in interstellar space is high because they are known or inferred to be abundant in cold dense clouds. Cyclopentadiene and its cyano- and ethynyl-derivatives (including the isomer of ethynyl cyclopentadiene fulvenallene) have been detected in \mbox{TMC-1} \citep{Cernicharo2021a,Cernicharo2021b,Cernicharo2022a,McCarthy2021,Lee2021a}, while in the case of benzene, its cyano and ethynyl derivatives have been detected as well \citep{McGuire2018,Loru2023} and there is ample evidence based on chemical kinetics \citep{Woon2006,Goulay2006,Balucani1999,Balucani2000,Landera2008,Trevitt2010,Jones2010,Cooke2020} that they are a proxy of the radio invisible molecule benzene. These mono-ringed structures are likely to open the door to form larger fused rings involving two or three cycles, several of which, such as indene, naphthalene, acenaphthylene, or pyrene, have been detected directly or through their cyano derivatives in \mbox{TMC-1} \citep{Cernicharo2021a,Cernicharo2024,Burkhardt2021a,McGuire2021,Wenzel2024}. Moreover, some of these aromatic cycles have been also detected in other cold dense clouds, in addition to \mbox{TMC-1} \citep{Burkhardt2021b,Agundez2023a}.

Several theoretical and experimental studies have provided evidence that 1,3-butadiene can be a precursor of cycles under interstellar conditions. For example, the reaction of 1,3-butadiene with C$_2$H is a promising route to form benzene \citep{Jones2011} and fulvene \citep{Lockyear2015}, while its reaction with CH can produce cyclopentadiene \citep{He2020}, which in turn can react with CH to produce benzene \citep{Caster2019,Caster2021}, and even 1,3-butadiene reacting with CH$_3$CC could form toluene \citep{Thomas2019}. The radical CH$_3$CC has not been for the moment observed in interstellar space but the radicals CH and C$_2$H are known to be abundant in cold dense clouds such as \mbox{TMC-1} \citep{Agundez2013}, making the aforementioned reactions promising routes to form cyclopentadiene and benzene.

The validation of the hypothesis that 1,3-butadiene is a key intermediate in the bottom-up synthesis of large cycles relies on whether it is abundant or not in cold dense clouds. 1,3-butadiene has several rotamers depending on the internal rotation about the single C-C bond. The most stable form is the \textit{s-trans}, while the \textit{s-gauche} and \textit{s-cis} lie 3.0 and 3.5 kcal mol$^{-1}$ (1500 and 1750 K), respectively, above the \textit{s-trans}, according to the theoretical calculations by \cite{Feller2009}. The \textit{s-trans} rotamer is non-polar and thus cannot detected at radio wavelengths. The next rotamers are expected to be polar but they have proven to be very elusive to experimental characterization.

Given that the lowest energy rotamer of 1,3-butadiene is non-polar, its detection must rely on indirect methods. One possibility is to search for its protonated form, as done for other non-polar molecules, such as N$_2$, CO$_2$, C$_3$, C$_5$, NCCN, and NC$_4$N \citep{Thaddeus1975,Minh1991,Pety2012,Cernicharo2022b,Agundez2015,Agundez2023b}. However, to our knowledge, the rotational spectrum of protonated 1,3-butadiene has not been measured in the laboratory, which prevents its radioastronomical search. An alternative indirect method is to search for the cyano substituted form. Indeed, several non-polar molecules, such as ethane, benzene, and naphtalene, are inferred to be present in cold dense clouds thanks to the detection of their cyano derivatives \citep{Cernicharo2021c,McGuire2018,McGuire2021}. The hypothesis that cyano derivatives can be used as a proxy of 1,3-butadiene is validated because experimental and theoretical studies show that 1,3-butadiene reacts rapidly with the radical CN at low temperatures \citep{Bullock1971,Butterfield1993,Morales2011,Gardez2012}, forming the polar molecule 1-cyano-1,3-butadiene \citep{Morales2011,Sun2014}, which can thus be searched provided their rotational spectra are known. 

\begin{figure*}
\centering
\includegraphics[angle=0,width=0.80\textwidth]{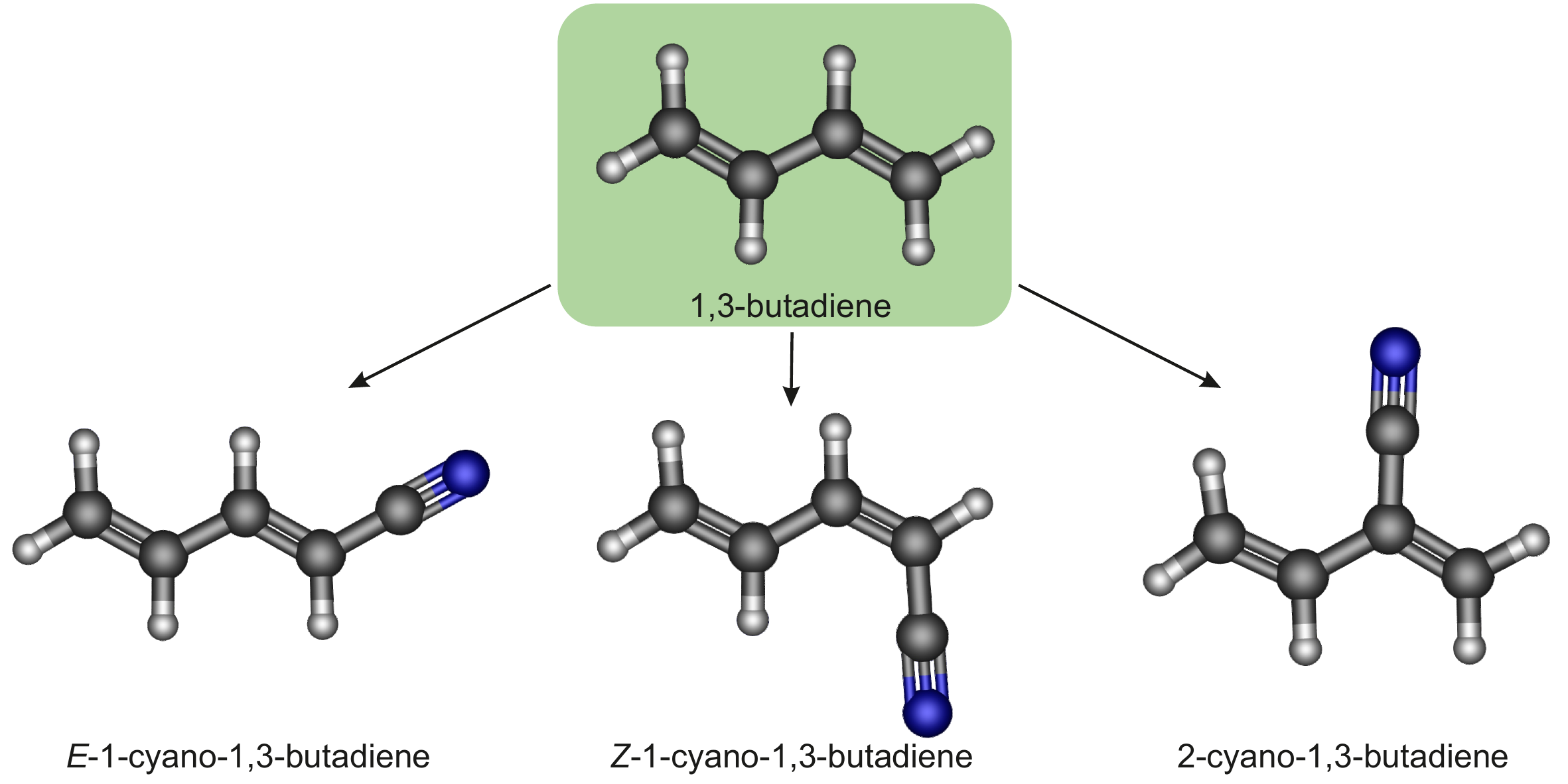}
\caption{Geometrical structure of the three cyano derivatives of 1,3-butadiene resulting from the substitution of one terminal H atom ($E$ and $Z$ isomers of 1-cyano-1,3-butadiene) and one internal H atom (2-cyano-1,3-butadiene).} \label{fig:cn-butadiene}
\end{figure*}

The rotational spectrum of 1-cyano-1,3-butadiene has been recently measured in the laboratory \citep{McCarthy2020,Zdanovskaia2021a}, while that of 2-cyano-1,3-butadiene has been measured as well \citep{Zdanovskaia2021a}. Based on these laboratory data, \cite{Cooke2023} found statistical evidence of $E$-1-cyano-1,3-butadiene in \mbox{TMC-1} using astronomical spectra obtained with the GBT telescope and a line stack technique. The statistical evidence of this isomer of cyano-1,3-butadiene is however modest, with a signal-to-noise ratio of 5.1\,$\sigma$. Given the importance of 1,3-butadiene as a potential precursor in bottom-up routes to synthesize large cycles in cold dense clouds, here we revisit the evidence for the presence of cyano-1,3-butadiene using independent data of \mbox{TMC-1} obtained with the Yebes\,40m telescope in the frame of the QUIJOTE (Q-band Ultrasensitive Inspection Journey to the Obscure TMC-1 Environment) project \citep{Cernicharo2021d}.

\section{Astronomical observations}

The astronomical observations presented here are part of the ongoing Yebes\,40m Q-band line survey of \mbox{TMC-1} QUIJOTE \citep{Cernicharo2021d}, which consists of a Q-band line survey (31.0-50.3 GHz) of the dark cloud \mbox{TMC-1} at the position of the cyanopolyyne peak ($\alpha_{J2000}=4^{\rm h} 41^{\rm  m} 41.9^{\rm s}$ and $\delta_{J2000}=+25^\circ 41' 27.0''$). The observations are carried out using the frequency-switching observing technique, with a frequency throw of either 8 or 10 MHz. The full Q band is covered in one shot with a spectral resolution of 38.15 kHz in both polarizations using a 7 mm dual linear polarization receiver connected to a set of 2\,$\times$\,8 fast Fourier transform spectrometer \citep{Tercero2021}. The intensity scale at the telescope is the antenna temperature, $T_A^*$, which has an estimated uncertainty due to calibration of 10\,\%, which can be converted to main beam brightness temperature, $T_{\rm mb}$, by dividing $T_A^*$ by $B_{\rm eff}$/$F_{\rm eff}$, where the beam efficiency $B_{\rm eff}$ is given by the Ruze formula $B_{\rm eff}$\,=\,0.797\,$\exp{[-(\nu/71.1)^2]}$, where $\nu$ is the frequency in GHz, and the forward efficiency $F_{\rm eff}$ is 0.97. The half power beam width (HPBW) can be approximated as HPBW($''$) = 1763/$\nu$(GHz). The latest dataset of QUIJOTE used here includes observations carried out between November 2019 and July 2024. The total on-source telescope time is 1509.2 h, of which 736.6 h correspond to a frequency throw of 8 MHz and 772.6 h to a throw of 10 MHz. The $T_A^*$ rms noise level varies between 0.06 mK at 32 GHz and 0.18 mK at 49.5 GHz. The procedure used to reduce and analyze the data is described in \cite{Cernicharo2022a}. The data were analyzed using the GILDAS software\footnote{\texttt{https://www.iram.fr/IRAMFR/GILDAS/}}.

\section{Spectroscopy of cyano-1,3-butadiene} \label{sec:spectroscopy}

The substitution of one hydrogen atom by a $-$CN group in 1,3-butadiene can result in different isomers (see Fig.\,\ref{fig:cn-butadiene}). If the substitution occurs in one of the four hydrogen atoms of the terminal carbon atoms, 1-cyano-1,3-butadiene is formed, while substitution in one of the two hydrogen atoms of the internal carbon atoms results in the 2-cyano-1,3-butadiene isomer. The most stable isomer is 1-cyano-1,3-butadiene, which in turn has $E$ and $Z$ isomers, the $E$ form being 0.3 kcal mol$^{-1}$ more stable than the $Z$ form \citep{Zdanovskaia2021a}. The isomer 2-cyano-1,3-butadiene is 1 kcal mol$^{-1}$ less stable than $E$-1-cyano-1,3-butadiene \citep{Sun2014}. These three isomers can exist in different conformations, the lowest in energy is $s$-$trans$ for the $E$ and $Z$ forms of 1-cyano-1,3-butadiene and $syn$ for 2-cyano-1,3-butadiene, all of them with sizable dipole moments of $\mu_a$\,=\,4.6\,D and $\mu_b$\,=\,1.0\,D for $E$-1-cyano-1,3-butadiene, $\mu_a$\,=\,3.6\,D and $\mu_b$\,=\,2.3\,D for $Z$-1-cyano-1,3-butadiene, and $\mu_a$\,=\,3.2\,D and $\mu_b$\,=\,2.3\,D for 2-cyano-1,3-butadiene \citep{Zdanovskaia2021a,Zdanovskaia2021b}.

\begin{figure*}
\centering
\includegraphics[angle=0,width=\textwidth]{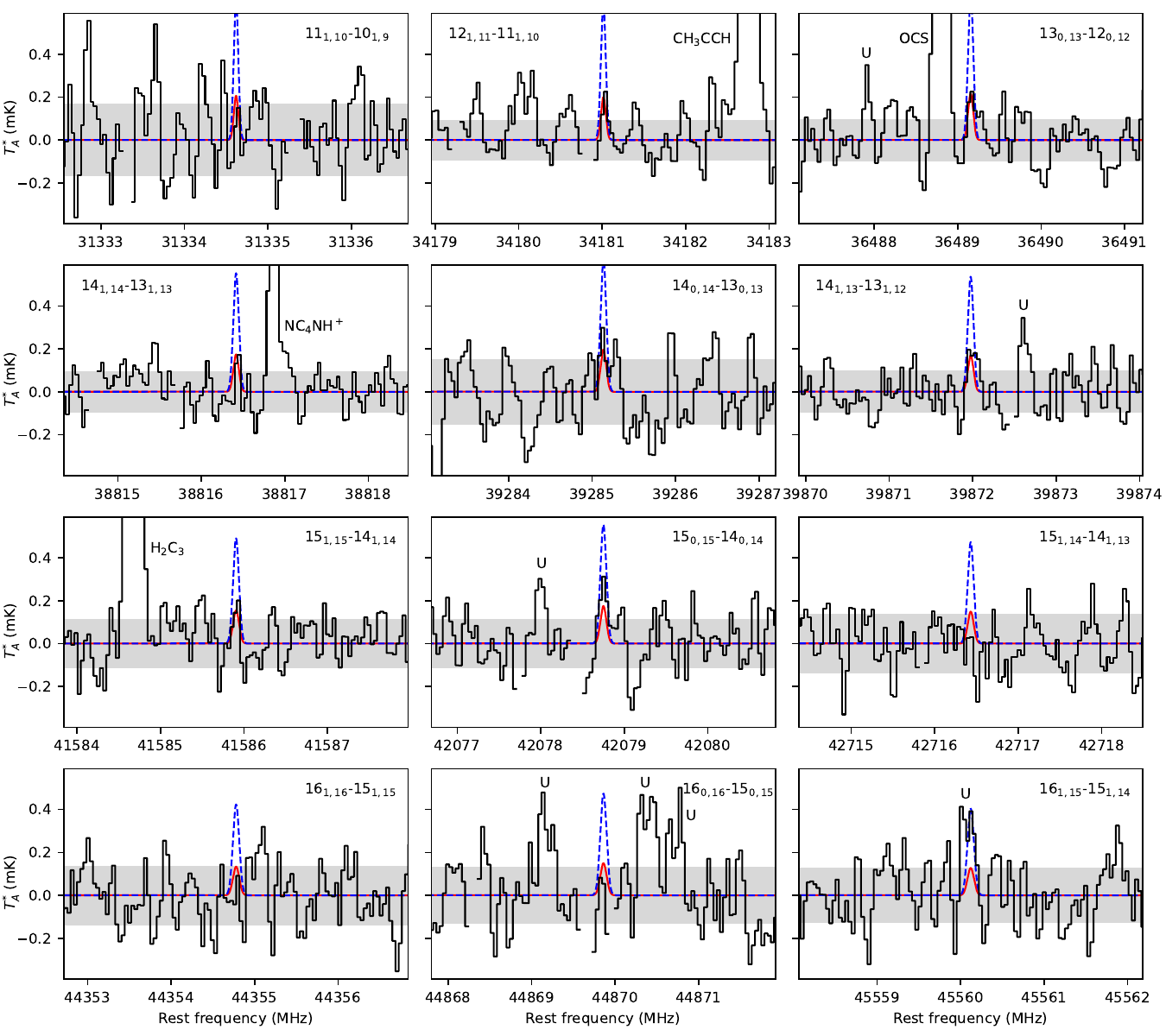}
\caption{Spectra of \mbox{TMC-1} in the Q band at the frequencies of the most favorable lines of $E$-1-cyano-1,3-butadiene. The noise level, as measured in a window of $\pm$\,8 MHz around the expected position of each line with the nominal spectral resolution of 38.15 kHz, is indicated by a gray horizontal band. The red lines correspond to the line intensities calculated adopting as column density for $E$-1-cyano-1,3-butadiene the 3\,$\sigma$ upper limit of 1.2\,$\times$\,10$^{10}$ cm$^{-2}$ determined here, while the blue dashed lines correspond to the line intensities calculated adopting as column density the value of 3.8\,$\times$\,10$^{10}$ cm$^{-2}$ determined by \cite{Cooke2023}. Our data is clearly inconsistent with such column density.} \label{fig:lines_e-ch2chchchcn}
\end{figure*}

The rotational spectra of the lowest $s$-$trans$ conformers of the $E$ and $Z$ isomers of 1-cyano-1,3-butadiene have been characterized in the laboratory in the frequency ranges 6.5-26 GHz \citep{McCarthy2020} and 130-375 GHz \citep{Zdanovskaia2021a}, while that of $syn$-2-cyano-1,3-butadiene has been measured in the laboratory by \cite{Zdanovskaia2021b} in the frequency range 130-360 GHz. Frequency predictions for the three cyano-1,3-butadiene isomers in the Q band were done using the molecular parameters derived from the rotational spectra analysis \citep{Zdanovskaia2021a,Zdanovskaia2021b} employing the SPCAT program \citep{Pickett1991}. In the case of the isomers of 1-cyano-1,3-butadiene, the availability of low-frequency data permitted to resolve the hyperfine structure due to the nitrogen nucleus, something that was not possible in the case of 2-cyano-1,3-butadiene. In this case, we estimated the $^{14}$N nuclear quadrupole coupling constants through quantum chemical calculations at the B3LYP/cc-pVTZ level of theory. We used this method because it reproduces quite well the experimental values of $\chi_{aa}$ and $\chi_{bb}$ derived for the $E$ and $Z$ isomers of 1-cyano-1,3-butadiene \citep{McCarthy2020,Zdanovskaia2021a}. For the $E$ isomer we calculate $\chi_{aa}$\,=\,$-$3.53 MHz and $\chi_{bb}$\,=\,1.40 MHz, in good agreement with the experimental values of $-$3.33 and 1.28 MHz, respectively, while for the $Z$ isomer we obtain $\chi_{aa}$\,=\,$-$0.69 MHz and $\chi_{bb}$\,=\,$-$1.40 MHz, in line with the experimental values of $-$0.36 and $-$1.65 MHz, respectively. For 2-cyano-1,3-butadiene we calculate $\chi_{aa}$\,=\,$-$1.29 MHz and $\chi_{bb}$\,=\,$-$0.89 MHz. For the three molecules, the frequency errors in the Q band are typically around 10-15 kHz for $a$-and $b$-type transitions with $K_a$\,=\,0, 1. The hyperfine splitting for these lines is modest, on the order of a few tens of kHz.

\section{Search for cyano-1,3-butadiene in \mbox{TMC-1}} \label{sec:results}

We first searched for the isomer $E$-1-cyano-1,3-butadiene in our Yebes\,40m data of \mbox{TMC-1} because \cite{Cooke2023} announced its detection based on a line stack analysis of GBT data of the same cloud. In order to identify the most favorable lines of this molecule in the Q band we computed the line intensities at a rotational temperature of 9 K, which is the gas kinetic temperature in \mbox{TMC-1} \citep{Agundez2023c}. The most intense predicted lines correspond to $a$-type transitions with $K_a$\,=\,0 and 1. We made a careful inspection of those lines in our QUIJOTE data. The 12$_{0,12}$-11$_{0,11}$ transition at 33690.962 MHz coincides with the $J$\,=\,13-12 line of HC$_5$NH$^+$ at 33690.975 MHz, which has a $T_A^*$ intensity of 9.7 mK \citep{Marcelino2020}, while there is a clear line with a $T_A^*$ intensity of 0.7 mK at the frequency of the 12$_{1,12}$-11$_{1,11}$ transition, which is 33275.732 MHz. This line however cannot be due to $E$-1-cyano-1,3-butadiene because many other lines should be present with similar intensities, which is not the case. We do not have a clear assignment for this line, which for the moment remains unidentified. There are still many lines with $T_A^*$ intensities below 1 mK in our QUIJOTE data that await identification.

\begin{figure*}
\centering
\includegraphics[angle=0,width=\textwidth]{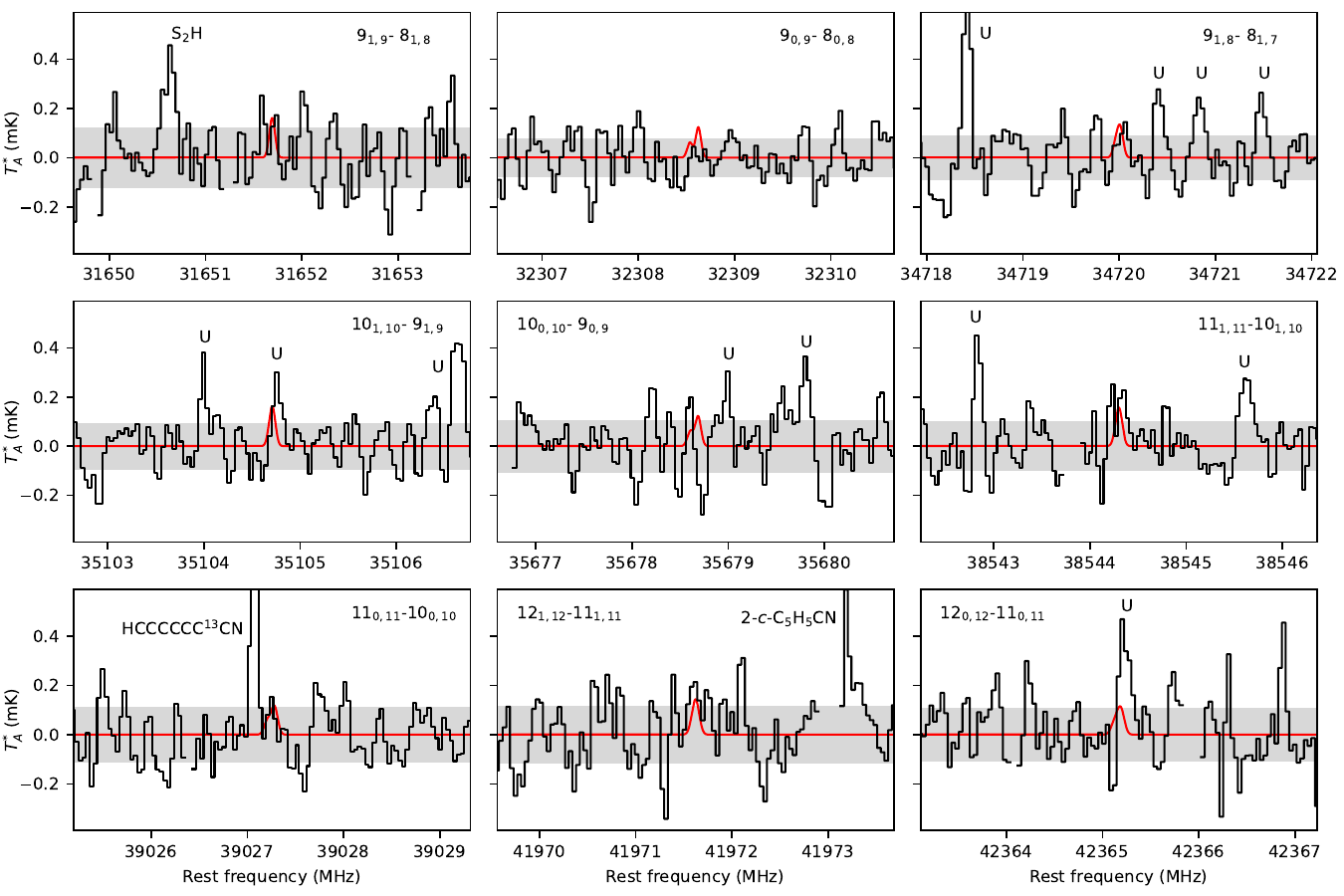}
\caption{Spectra of \mbox{TMC-1} in the Q band at the frequencies of the most favorable lines of $Z$-1-cyano-1,3-butadiene. The noise level, as measured in a window of $\pm$\,8 MHz around the expected position of each line with the nominal spectral resolution of 38.15 kHz, is indicated by a gray horizontal band. The red lines correspond to the line intensities calculated adopting as column density for $Z$-1-cyano-1,3-butadiene the 3\,$\sigma$ upper limit of 2.0\,$\times$\,10$^{10}$ cm$^{-2}$ determined here.} \label{fig:lines_z-ch2chchchcn}
\end{figure*}

In Figure\,\ref{fig:lines_e-ch2chchchcn} we show the spectra at the expected position of the most favorable lines of $E$-1-cyano-1,3-butadiene that lie in low-noise and relatively clean spectral regions. At the position of the 15$_{0,15}$-14$_{0,14}$ and 16$_{1,15}$-15$_{1,14}$ transitions the data show emission features with $T_A^*$ intensities of 0.3 mK and 0.4 mK, respectively, which are probably real lines. However, their assignment to $E$-1-cyano-1,3-butadiene is inconsistent with the non-detection of other lines that should have comparable or even higher intensities. As for the rest of transitions shown in Fig.\,\ref{fig:lines_e-ch2chchchcn}, for some of them there is a feature at the expected position of the line but it lies within the noise level of the spectrum. We must conclude that we do not have enough evidence for the presence of $E$-1-cyano-1,3-butadiene in \mbox{TMC-1}. The non-detection of the 13$_{0,13}$-12$_{0,12}$ line at 36489.156 MHz provides the most stringent upper limit to the column density of this isomer of cyano-1,3-butadiene. Assuming a rotational temperature of 9 K and a line width of 0.6 km s$^{-1}$ \citep{Agundez2023c}, we derive a 3\,$\sigma$ upper limit to the beam-averaged column density of $E$-1-cyano-1,3-butadiene in \mbox{TMC-1} of 1.2\,$\times$\,10$^{10}$ cm$^{-2}$, which is 3.2 times lower than the column density determined by \cite{Cooke2023}. The calculated line profiles adopting as column density the 3\,$\sigma$ upper limit derived here are shown in red in Fig.\,\ref{fig:lines_e-ch2chchchcn}. If the column density of $E$-1-cyano-1,3-butadiene were 3.8\,$\times$\,10$^{10}$ cm$^{-2}$, as derived by \cite{Cooke2023}, the lines of this molecule should be clearly visible above the noise level in our sensitive Q band data (see blue dashed lines in Fig.\,\ref{fig:lines_e-ch2chchchcn}). The calculated the line profiles assume that the emission size of $E$-1-cyano-1,3-butadiene would be extended compared to the main beam of the Yebes\,40m (35-57\,$''$ in the Q band) and GBT (20-94\,$''$ in the X, K and Ka bands) telescopes. Although we do not have information on the true emission size of $E$-1-cyano-1,3-butadiene, if this species is indeed present in \mbox{TMC-1}, this is a reasonable assumption given the extended nature of the emission observed for several hydrocarbons and N-bearing molecules in \mbox{TMC-1} \citep{Pratap1997,Fosse2001,Cernicharo2023}.

We also searched for the $Z$ isomer of 1-cyano-1,3-butadiene in our \mbox{TMC-1} data. As with the $E$ isomer, for a rotational temperature of 9 K, the most favorable lines for its detection in the Q band correspond to $a$-type transitions with $K_a$\,=\,0 and 1. In Fig.\,\ref{fig:lines_z-ch2chchchcn} we show spectra at the frequencies of some of these favorable lines that lie in uncontaminated and sensitive spectral regions. As in the case of the $E$ isomer, for some of the transitions there is an emission feature around the expected position of the line. For example, in the case of the 10$_{1,10}$-9$_{1,9}$ and 12$_{0,12}$-11$_{0,11}$ transitions, there are emission features with $T_A^*$ intensities of 0.3 mK and 0.5 mK, respectively, which are very likely real lines although they cannot be assigned to $Z$-1-cyano-1,3-butadiene because on the one hand they appear slightly shifted with respect to the expected position and on the other their intensities are inconsistent with the non-detection of other lines that should have comparable of even larger intensities. Here we conclude that $Z$-1-cyano-1,3-butadiene is not detected and derive a 3\,$\sigma$ upper limit to its beam-averaged column density of 2.0\,$\times$\,10$^{10}$ cm$^{-2}$.

\begin{figure*}
\centering
\includegraphics[angle=0,width=\textwidth]{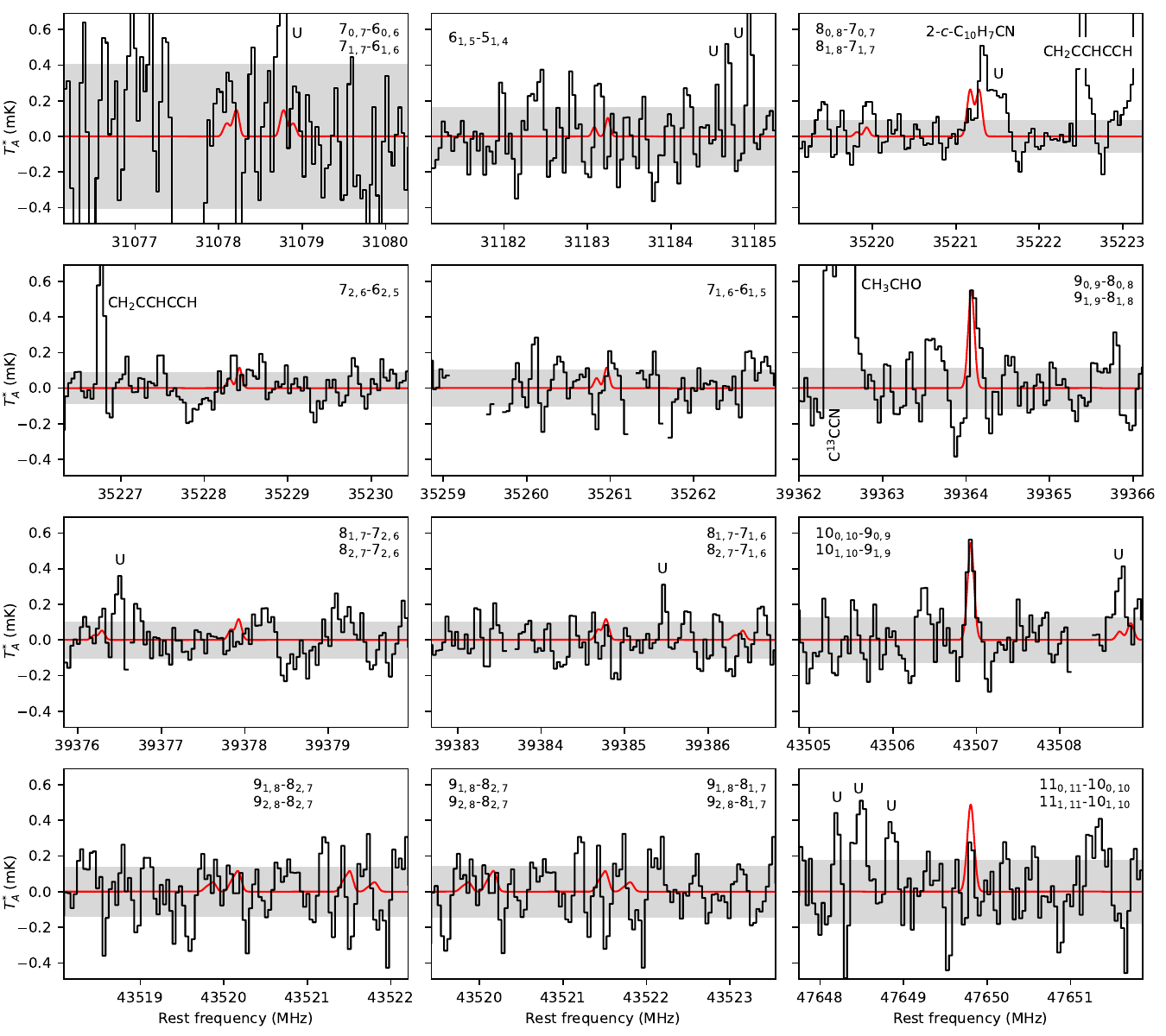}
\caption{Spectra of \mbox{TMC-1} in the Q band at the frequencies of the most favorable lines of 2-cyano-1,3-butadiene. The noise level, as measured in a window of $\pm$\,8 MHz around the expected position of each line with the nominal spectral resolution of 38.15 kHz, is indicated by a gray horizontal band. The red lines correspond to the line intensities calculated adopting a column density of 2-cyano-1,3-butadiene of 3.1\,$\times$\,10$^{10}$ cm$^{-2}$.} \label{fig:lines_ch2chccnch2}
\end{figure*}

The astronomical search for the third isomer of cyano-1,3-butadiene, which is 2-cyano-1,3-butadiene, was carried out similarly to that of the two other isomers. Again, the most favorable lines correspond to rotational transitions of $a$-type with quantum number $K_a$\,=\,0 and 1. In this particular case, the transitions $J$$_{0,J}$\,$\rightarrow$\,$J$$-$1$_{0,J-1}$ and $J$$_{1,J}$\,$\rightarrow$\,$J$$-$1$_{1,J-1}$, which are predicted to be the brightest, tend to collapse to the same frequency as the quantum number $J$ increases. Therefore, these pairs of lines stand out as the most favorable ones to search for 2-cyano-1,3-butadiene in the Q band. The transitions 7$_{0,7}$-6$_{0,6}$ and 7$_{1,7}$-6$_{1,6}$ are separated by 566 kHz and lie in a relatively noisy region of the spectrum (top-left panel in Fig.\,\ref{fig:lines_ch2chccnch2}), while the transitions 8$_{0,8}$-7$_{0,7}$ and 8$_{1,8}$-7$_{1,7}$, which are separated by 94 kHz, coincide with a weak emission feature at the noise level seen as a shoulder in the red side of a $T_A^*$\,$\sim$\,0.5 mK line (top-right panel in Fig.\,\ref{fig:lines_ch2chccnch2}). The next two pairs of lines with $J$\,=\,9 and $J$\,=\,10 are separated by less than 15 kHz, which is smaller than the spectral resolution of 38.15 kHz of our \mbox{TMC-1} data, and coincide with lines of $T_A^*$ intensities of 0.5 mK that are seen above the noise level (two middle-right panels in Fig.\,\ref{fig:lines_ch2chccnch2}). The next pair of lines, 11$_{0,11}$-10$_{0,10}$ and 11$_{1,11}$-10$_{1,10}$, lies again in a relatively noisy spectrum and no clear feature can be appreciated above the noise level (bottom-right panel in Fig.\,\ref{fig:lines_ch2chccnch2}). Other transitions with $K_a$\,=\,1 and 2 are not clearly seen at the current level of sensitivity of our data (remaining panels in Fig.\,\ref{fig:lines_ch2chccnch2}). Therefore, the current evidence for the presence of 2-cyano-1,3-butadiene in \mbox{TMC-1} comes from two lines that coincide with the pairs 9$_{0,9}$-8$_{0,8}$/9$_{1,9}$-8$_{1,8}$ at 39364.05 MHz and 10$_{0,10}$-9$_{0,9}$/10$_{1,10}$-9$_{1,9}$ at 43506.93 MHz. With just two lines and given the relatively high density of lines with $T_A^*$ intensities below 1 mK, we must be cautious. We thus conclude that for the moment there is tentative evidence for the presence of 2-cyano-1,3-butadiene in \mbox{TMC-1}. We derive an upper limit to its column density of 3.1\,$\times$\,10$^{10}$ cm$^{-2}$. The confirmation of this detection must await more sensitive data in the Q band. We note that, unlike in the cases of the $E$ and $Z$ isomers of 1-cyano-1,3-butadiene, the hyperfine splitting of the rotational transitions of 2-cyano-1,3-butadiene is somewhat uncertain because it is based on ab initio calculations and not in experimental measurements. The impact of the hyperfine splitting on the tentative detection is however likely to be small because the predicted splitting, which is similar to the experimentally determined ones for the two isomers of 1-cyano-1,3-butadiene, is comparable or smaller than the spectral resolution of the \mbox{TMC-1} data.

\section{Discussion} \label{sec:discussion}

\begin{figure}
\centering
\includegraphics[angle=0,width=\columnwidth]{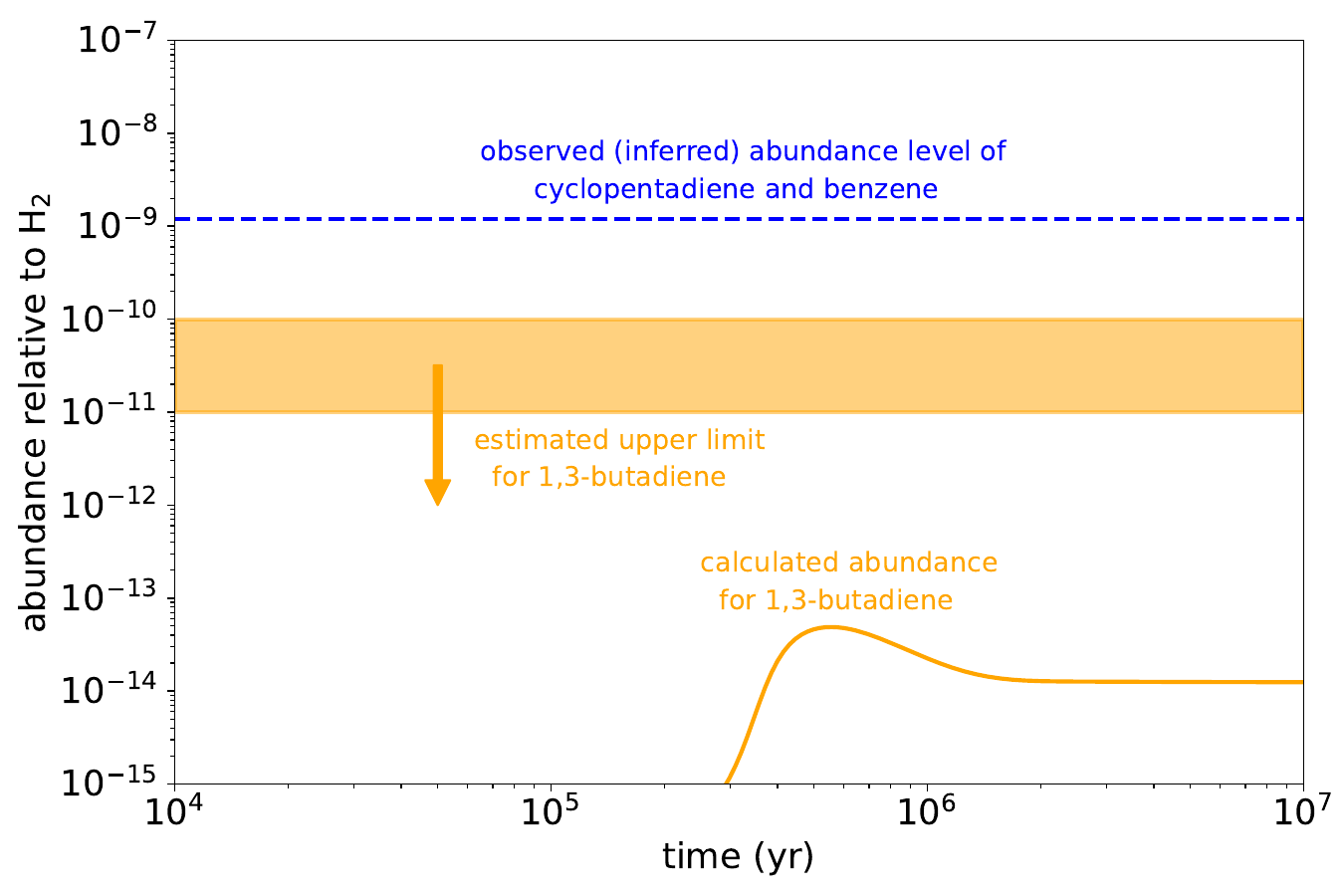}
\caption{Abundance of 1,3-butadiene calculated by the chemical model as a function of time (solid orange curve), estimated observational upper limit on the abundance of 1,3-butadiene in \mbox{TMC-1} (horizontal orange band), and abundance observed (or inferred) for cyclopentadiene and benzene in \mbox{TMC-1}.} \label{fig:abun}
\end{figure}

The non-detection of 1-cyano-1,3-butadiene and the tentative evidence of 2-cyano-1,3-butadiene can shed light on the abundance of 1,3-butadiene in \mbox{TMC-1} and its potential role as intermediate in the synthesis of simple mono-ringed cyclic structures such as cyclopentadiene and benzene. To this regard it is interesting to have a look at the typical abundance ratios between a given molecule and its cyano derivative in \mbox{TMC-1}. There are various cases in which this abundance ratio has been observationally constrained. For example, the three C$_4$H$_3$N isomers CH$_3$C$_3$N, CH$_2$CCHCN, and HCCCH$_2$CN are probably produced in the reactions of CN with CH$_3$CCH and CH$_2$CCH$_2$ \citep{Carty2001,Balucani2000,Balucani2002,Abeysekera2015}. These three cyano derivatives together are 18 times less abundant than CH$_3$CCH, which is the polar hydrocarbon precursor detectable in \mbox{TMC-1} \citep{Marcelino2021,Cernicharo2022a}. The abundance of propene is 90 times higher than the sum of the abundances of the five cyano derivatives detected in \mbox{TMC-1} \citep{Marcelino2007,Cernicharo2022c}, which are likely formed in the reaction between propene and CN \citep{Morales2010,Gu2008,Huang2009}. It is also known that maleonitrile (NC$-$CH$=$CH$-$CN) is 130 times less abundant than CH$_2$CHCN \citep{Gratier2016,Agundez2024}, its likely precursor via the reaction with CN \citep{Marchione2022}. For other cyano derivatives observed in \mbox{TMC-1}, there is no experimental or theoretical evidence that the reaction with CN is the main source of them, although based on similar reactions it is a likely possibility. These cases comprise the cyano derivatives of vinyl acetylene, CHCCHCHCN and CH$_2$CHC$_3$N, which together account for 1/25 of the abundance of vinyl acetylene \citep{Cernicharo2021c,Lee2021b}, the cyanide CH$_2$CCHC$_3$N, which has an abundance 1/60 times that of CH$_2$CCHCCH \citep{Shingledecker2021,Cernicharo2021e}, the cyano derivatives of cyclopentadiene, which are 30 times less abundant than cyclopentadiene \citep{Cernicharo2021a,Cernicharo2021b}, and the cyano derivative of indene, which is present with an abundance 1/80 times of that of indene \citep{Cernicharo2021a,Sita2022}. It is also worth to note that the methyl cyanopolyynes CH$_3$C$_5$N and CH$_3$C$_5$N are 8-9 times less abundant than their potential precursors via reaction with CN, CH$_3$C$_4$H and CH$_3$C$_6$H, respectively \citep{Agundez2013,Siebert2022,Cernicharo2022a}. In summary, the abundance ratios found in \mbox{TMC-1} between a given precursor (via reaction with CN) and the resulting cyano derivative are in the range 8-130.

Based on the above observational information, we could estimate that 1,3-butadiene is present in \mbox{TMC-1} with an abundance 8-130 times higher than that of 1-cyano-1,3-butadiene, which is the main product of the reaction between 1,3-butadiene and CN, according to crossed molecular beam experiments and theoretical calculations \citep{Morales2011,Sun2014}. Chemical model calculations can also provide some insights on the abundance ratio between 1,3-butadiene and 1-cyano-1,3-butadiene. In their chemical model, \cite{Cooke2023} predict a value of $\sim$\,10 for this abundance ratio assuming that 1-cyano-1,3-butadiene is mostly destroyed through reactions with cations. We note that the abundance ratio of $\sim$\,10 would increase if 1-cyano-1,3-butadiene is assumed to react rapidly with neutral atoms, such as C and H. In any case, it is encouraging that the findings from chemical models are consistent with the range of abundance ratios derived from observations. Since here we find that 1-cyano-1,3-butadiene is present with an abundance below 1.2\,$\times$\,10$^{-12}$ relative to H$_2$, adopting the upper limit on the column density of $E$-1-cyano-1,3-butadiene derived here and a column density of H$_2$ of 10$^{22}$ cm$^{-2}$ \citep{Cernicharo1987}, we can conclude that the abundance of 1,3-butadiene in \mbox{TMC-1} is below 8-130 times that of 1-cyano-1,3-butadiene, i.e., approximately below 10$^{-11}$-10$^{-10}$ with respect to H$_2$ (see Fig.\,\ref{fig:abun}).

To shed further light on the abundance of 1,3-butadiene in cold dense clouds we carried out chemical modeling calculations using the latest release of the UMIST chemical network \citep{Millar2024}, which includes a reduced set of reactions involving 1,3-butadiene. We adopted physical conditions and elemental abundances typical of cold dense clouds, with the abundance of oxygen depleted to result in a C/O elemental abundance ratio of 1  \citep{Agundez2013}. There is evidence that oxygen could be depleted on dust grains in cold dense clouds \citep{Jenkins2009,Whittet2010,Hincelin2011}. This would lead to an increase in the gas-phase elemental C/O ratio and enhance the abundances of C-bearing molecules \citep{Agundez2013}. The calculated abundance of 1,3-butadiene is rather low, it lies below 10$^{-13}$ relative to H$_2$ at any time (see Fig.\,\ref{fig:abun}). According to the chemical model, 1,3-butadiene is mainly formed by the dissociative recombination with electrons of the cation C$_4$H$_7^+$, which arises mainly from the reaction C$_2$H$_4^+$ + C$_2$H$_4$, while it is mostly destroyed by reactions with neutral atoms, such as C and H, and abundant cations, such as HCO$^+$. This chemical scheme reveals as insufficient to produce 1,3-butadiene with a high abundance. The non-detection of the two isomers of 1-cyano-1,3-butadiene is in line with a low abundance of 1,3-butadiene in \mbox{TMC-1}. The chemical model of \cite{Cooke2023} predicts an abundance of 1,3-butadiene somewhat higher than our model, although their maximum calculated abundance is also relatively low, a few times 10$^{-12}$ relative to H$_2$, which is within the conservative observational upper limit of $<$\,10$^{-11}$-10$^{-10}$ relative to H$_2$ estimated here.

Given that 1,3-butadiene seems to be less abundant than 10$^{-11}$-10$^{-10}$ relative to H$_2$, it is unlikely that it plays a key role as precursor, via reactions with CH and C$_2$H \citep{He2020,Jones2011}, of cyclopentadiene and benzene, which are observed (or inferred in the case of benzene) to be present with rather large abundances, around 10$^{-9}$ relative to H$_2$ (\citealt{Cernicharo2021a,Cernicharo2021b,Cernicharo2022a}; see Fig.\,\ref{fig:abun}). In cold dense clouds, a precursor of a given molecule may be less abundant than the molecule itself, but this usually happens when the precursor is an unstable species, such as a radical or an ion. This is a general behavior predicted by chemical models, and in some cases also confirmed by observations. For example, in \mbox{TMC-1} the cation H$_2$CCCH$^+$ is observed with a column density of 7.0\,$\times$\,10$^{11}$ cm$^{-2}$ \citep{Silva2023} and it is thought to be the precursor of H$_2$C$_3$, which is found to be around three times more abundant \citep{Cernicharo2022a}. In this case, the precursor 1,3-butadiene is a stable molecule, and therefore, any species formed from it is unlikely to be more abundant. Note that, as discussed above, stable precursors reacting with CN produce cyano derivatives that are systematically less abundant. Although in each particular case, the relative abundances between a precursor and a product will depend on the particular reactivities of each of them, we find that given the stable nature of both the precursor 1,3-butadiene and the products cyclopentadiene and benzene, it is unlikely that 1,3-butadiene can act as precursor if it is less abundant than the putative products.

The non-detection of 1-cyano-1,3-butadiene implies that 1,3-butadiene is probably not an important precursor in the synthesis of the first mono-ringed cycles. However, if the tentative evidence for 2-cyano-1,3-butadiene is confirmed at some point in the future, it is not straightforward how could this isomer be formed. Given that experiments and theoretical calculations exclude that this isomer can be formed through the reaction of CN and 1,3-butadiene \citep{Morales2011,Sun2014}, other neutral-neutral reactions that could a priori form CH$_2$C(CN)CHCH$_2$ are
\begin{equation}
\rm C_2H_3 + CH_2CHCN \rightarrow CH_2C(CN)CHCH_2 + H, \label{reac:c2h3+ch2chcn}
\end{equation}
\begin{equation}
\rm CH_2CCN + C_2H_4 \rightarrow CH_2C(CN)CHCH_2 + H. \label{reac:ch2ccn+c2h4}
\end{equation}
For these reactions to provide efficient routes to 2-cyano-1,3-butadiene in \mbox{TMC-1}, the reactants must be abundant enough and the reaction must be rapid and produce CH$_2$C(CN)CHCH$_2$. Given the relatively low abundance derived for 2-cyano-1,3-butadiene in \mbox{TMC-1}, $<$\,3.1\,$\times$\,10$^{-12}$ relative to H$_2$, the reactants involved in the two above reactions are probably abundant enough. CH$_2$CHCN and CH$_2$CCN have been detected in \mbox{TMC-1} with abundances of 6.5\,$\times$\,10$^{-10}$ relative to H$_2$ \citep{Gratier2016} and 2.5\,$\times$\,10$^{-11}$ \citep{Cabezas2023}, while the vinyl radical (C$_2$H$_3$) and ethylene (C$_2$H$_4$) are predicted by our chemical model to reach peak abundances relative to H$_2$ of $\sim$\,10$^{-9}$ and $\sim$\,10$^{-8}$, respectively. To our knowledge, the kinetics and product distribution of reactions (\ref{reac:c2h3+ch2chcn}) and (\ref{reac:ch2ccn+c2h4}) have not been investigated neither experimentally nor theoretically, and thus it is difficult to anticipate their expected behavior. Addition of vinyl radicals to multiple C$-$C bonds in hydrocarbons tends to have barriers \citep{Ismail2007,Smith2020}, although it is unknown whether reaction (\ref{reac:c2h3+ch2chcn}) would show a similar behavior. The kinetics of the radical CH$_2$CCN involved in reaction (\ref{reac:ch2ccn+c2h4}) is completely unexplored. Further studies on reactions (\ref{reac:c2h3+ch2chcn}) and (\ref{reac:ch2ccn+c2h4}) could shed light on their potential role in the synthesis of 2-cyano-1,3-butadiene.

\section{Conclusions}

We searched for the three isomers of cyano-1,3-butadiene in the cold dense cloud \mbox{TMC-1} using the latest QUIJOTE data obtained with the Yebes\,40m telescope. We do not find evidence for the $E$ or $Z$ isomers of 1-cyano-1,3-butadiene, and derive 3\,$\sigma$ upper limits to their column densities of 1.2\,$\times$\,10$^{10}$ cm$^{-2}$ and 2.0\,$\times$\,10$^{10}$ cm$^{-2}$, respectively. Our 3\,$\sigma$ upper limit on the column density of $E$-1-cyano-1,3-butadiene is 3.2 times lower than the column density determined by \cite{Cooke2023} using a line stack technique. We find tentative evidence for the presence of the third isomer, 2-cyano-1,3-butadiene, and derive an upper limit to its column density of 3.1\,$\times$\,10$^{10}$ cm$^{-2}$. Since this isomer is unlikely to result from the reaction between CN and 1,3-butadiene, we suggest that it could arise from neutral-neutral reactions, such as C$_2$H$_3$ + CH$_2$CHCN and CH$_2$CCN + C$_2$H$_4$. The upper limit imposed on the abundance of 1-cyano-1,3-butadiene allows us to estimate that the abundance of 1,3-butadiene in \mbox{TMC-1} is below 10$^{-11}$-10$^{-10}$ relative to H$_2$. This low abundance makes it unlikely that 1,3-butadiene plays an important role in bottom-up routes to the first mono-ringed cycles cyclopentadiene and benzene in cold dense clouds such as \mbox{TMC-1}. Therefore, other routes must be explored.

\begin{acknowledgements}

We acknowledge funding support from Spanish Ministerio de Ciencia, Innovaci\'on, y Universidades through grants PID2022-137980NB-I00 and PID2023-147545NB-I00.

\end{acknowledgements}

\end{document}